\documentclass[pra,a4paper,onecolumn,]{revtex4}
\usepackage[T1]{fontenc}
\usepackage{ae}
\usepackage{geometry}
\usepackage{comment}
\usepackage{amsmath}
\usepackage{graphicx}
\usepackage{enumerate}
\usepackage{subfigure}
\usepackage{multirow}
    
\begin{document}

\title{Observation of a quantum Cheshire Cat in a matter wave interferometer experiment}
 \author{Tobias Denkmayr$^1$}
 \author{Hermann Geppert$^1$}
 \author{Stephan Sponar$^1$}
\author{Hartmut Lemmel$^{1,2}$}
\author{Alexandre Matzkin$^3$}
\author{Jeff Tollaksen$^4$}
\author{Yuji Hasegawa$^1$}
\email{Hasegawa@ati.ac.at}
 \affiliation{%
$^1$Atominstitut, Vienna University of Technology,
Stadionallee 2, 1020 Vienna, Austria\\$^2$Institut Laue-Langevin, 6, Rue Jules Horowitz, 38042 Grenoble Cedex 9, France\\$^3$ Laboratoire de Physique Th\'{e}orique et Mod\'{e}lisation, CNRS Unit\'{e} 8089, Universit\'{e} de Cergy-Pontoise, 95302 Cergy-Pontoise cedex, France0\\$^4$Institute for Quantum Studies and Schmid School of Science and Technology, Chapman University, One University Drive, Orange, CA 92866, USA}

\date{\today}

\maketitle

\textbf{From its very beginning quantum theory has been revealing extraordinary and counter-intuitive phenomena, such as wave-particle duality \cite{bohreinstein51}, Schr\"odinger cats \cite{schrodinger35} and quantum non-locality \cite{EPR35,bohm51,bell64,bell66}. In the study of quantum measurement, a process involving pre- and postselection of quantum ensembles in combination with a weak interaction was found to yield unexpected outcomes \cite{aav88}. This scheme, usually referred to as "weak measurements", can not only be used as an amplification technique \cite{ritchie91,hosten08,dixon09} and for minimal disturbing measurements \cite{palacios-laloy10,steinberg11}, but also for the exploration of quantum paradoxes \cite{AV91,hardy92,aharonov05,lundeen09,yokota09}. Recently the quantum Cheshire Cat has attracted attention \cite{tollaksen01,aharonov13,matzkin13}: a quantum system can behave as if a particle and its property (e.g. its polarization) are spatially separated. Up to now most experiments studying weak measurements were done with photonic setups \cite{kofman12}.
To reveal the peculiarities of a quantum Cheshire Cat the use of non-zero mass particles is most appealing, since no classical description is possible. Here, we report an experiment using a neutron interferometer \cite{rauch00,rauch02,hasegawa03,hasegawa11} to create and observe a purely quantum mechanical Cheshire Cat. The experimental results suggest that the system behaves as if the neutrons went through one beam path, while their spin travelled along the other.}\\

\begin{figure}[!t]
\includegraphics[width=1\textwidth]{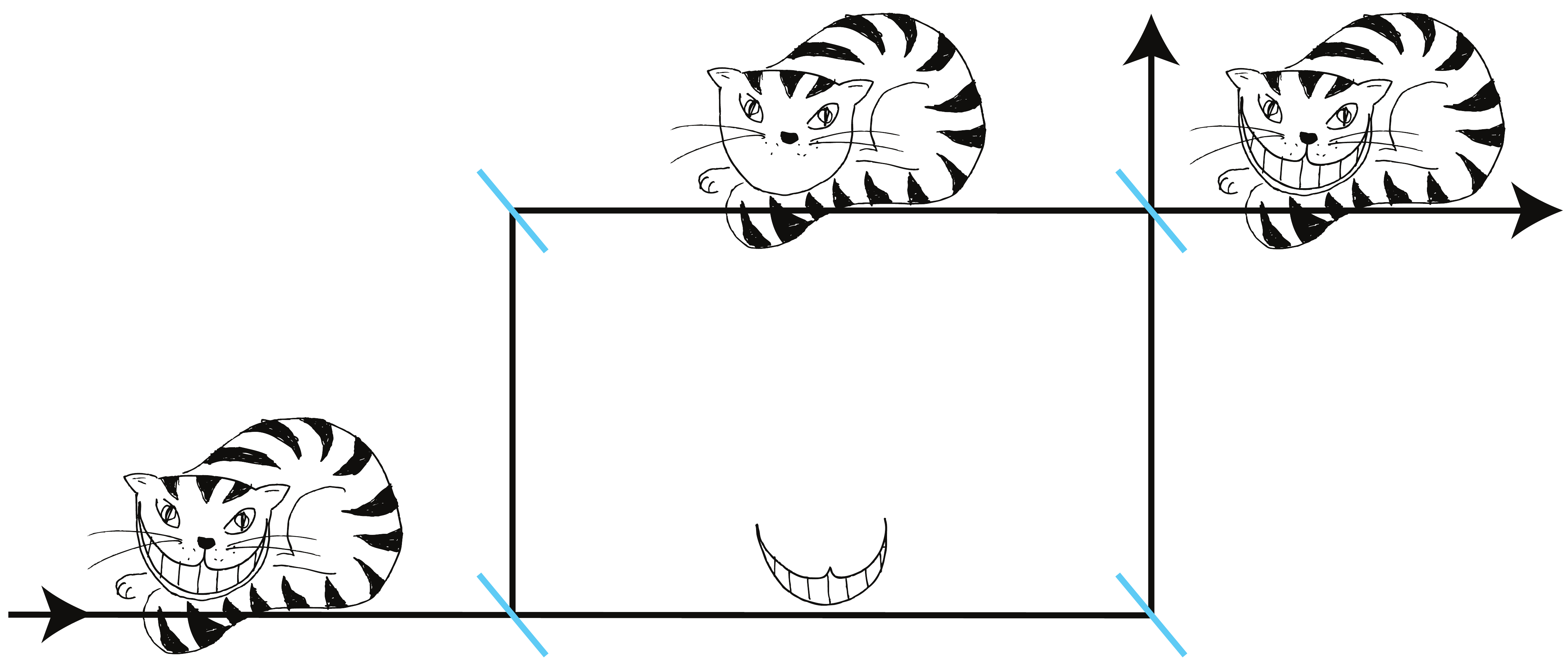}
\caption{Artistic depiction of the quantum Cheshire Cat: Inside the interferometer the Cat goes through the upper beam path, while its grin travels along the lower beam path.}\label{fig:cat}
\end{figure}

\begin{figure}[!t]
\includegraphics[width=1\textwidth]{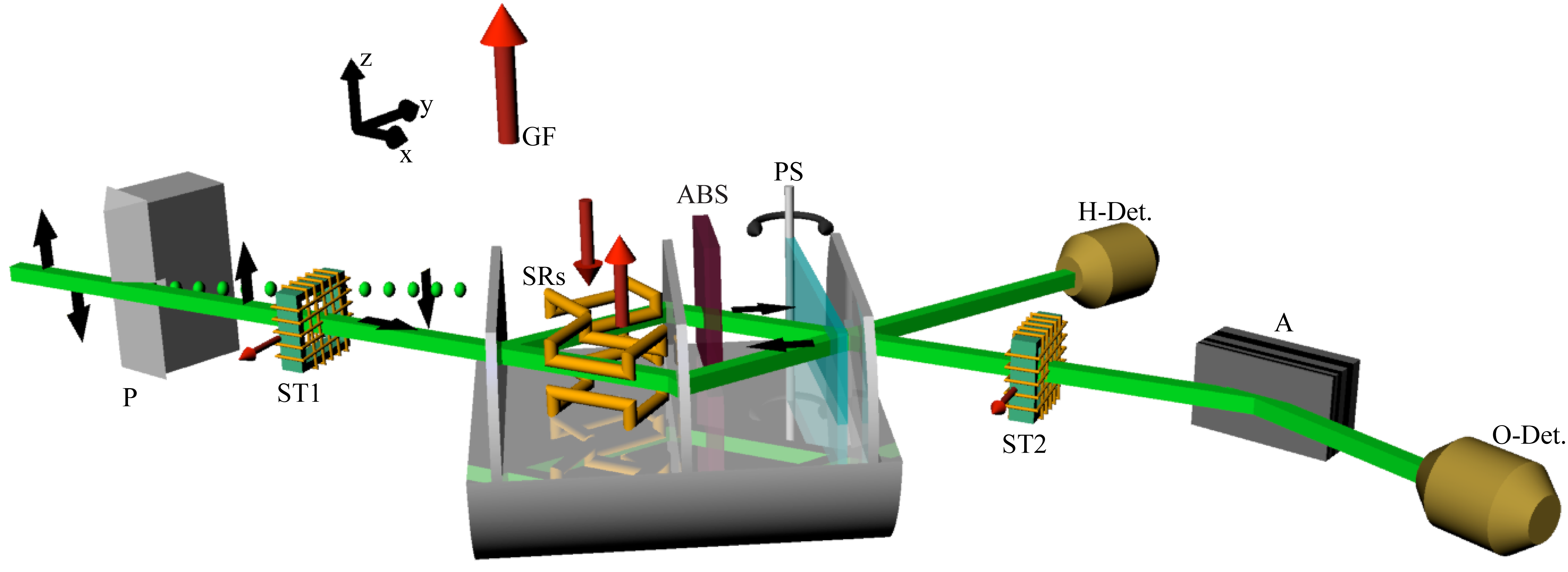}
\caption{Illustration of the experimental setup for the observation of a quantum Cheshire Cat in a neutron interferometer: The neutron beam is polarized by passing through magnetic birefringent prisms (P). To prevent depolarization, a magnetic guide field (GF) is applied around the whole setup. A spin turner (ST1) rotates the neutron spin by $\pi/2$. When entering the interferometer the neutron beam splits into two paths. Preselection of the system's wave function $|\psi_i\rangle$ is completed by two spin rotators (SRs) inside the neutron interferometer. These SRs are also used to perform the weak measurement of $\langle \hat{\sigma}_z\hat{\Pi}_{{I}}\rangle_w$ and $\langle \hat{\sigma}_z\hat{\Pi}_{{II}}\rangle_w$. The absorbers (ABS) are inserted in the beam paths when $\langle\hat{\Pi}_{{I}}\rangle_w$ and $\langle \hat{\Pi}_{{II}}\rangle_w$ are  determined. The phase shifter (PS) makes it possible to tune the relative phase $\chi$ between the beams in path $I$ and path $II$. The two outgoing beams of the interferometer are monitored by the H- and O-detector in reflected and forward directions, respectively. Only the neutrons reaching the O-detector are affected by postselection using a spin turner (ST2) and a spin analyzer (A).}\label{fig:setup}
\end{figure}

\begin{figure}[!t]
\includegraphics[width=1\textwidth]{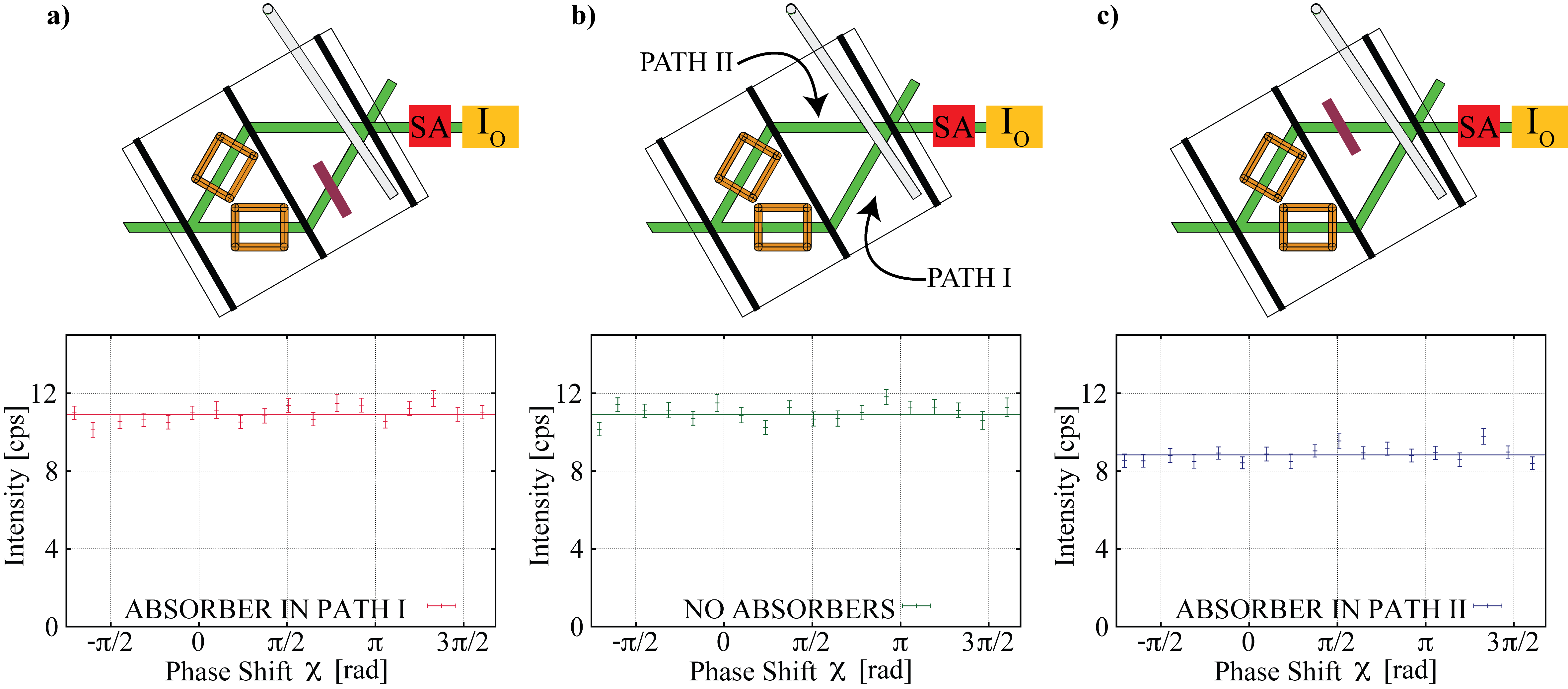}
\caption{Measurement of $\langle\hat{\Pi}_{{I}}\rangle_w$ and $\langle\hat{\Pi}_{{II}}\rangle_w$ using an absorber with the transmissivity $T=0.79(1)$. The intensity is plotted as a function of the relative phase $\chi$. \textbf{a)} An absorber in path $I$: No loss in intensity is recorded.  \textbf{b)} A reference measurement without any absorber \textbf{c)} An absorber in path $II$: The intensity decreases. Theses results suggest that for the successfully postselected ensemble the neutrons' go through path $II$.}\label{fig:setup_allabs_meas}
\end{figure}

\begin{figure}[!t]
\includegraphics[width=1\textwidth]{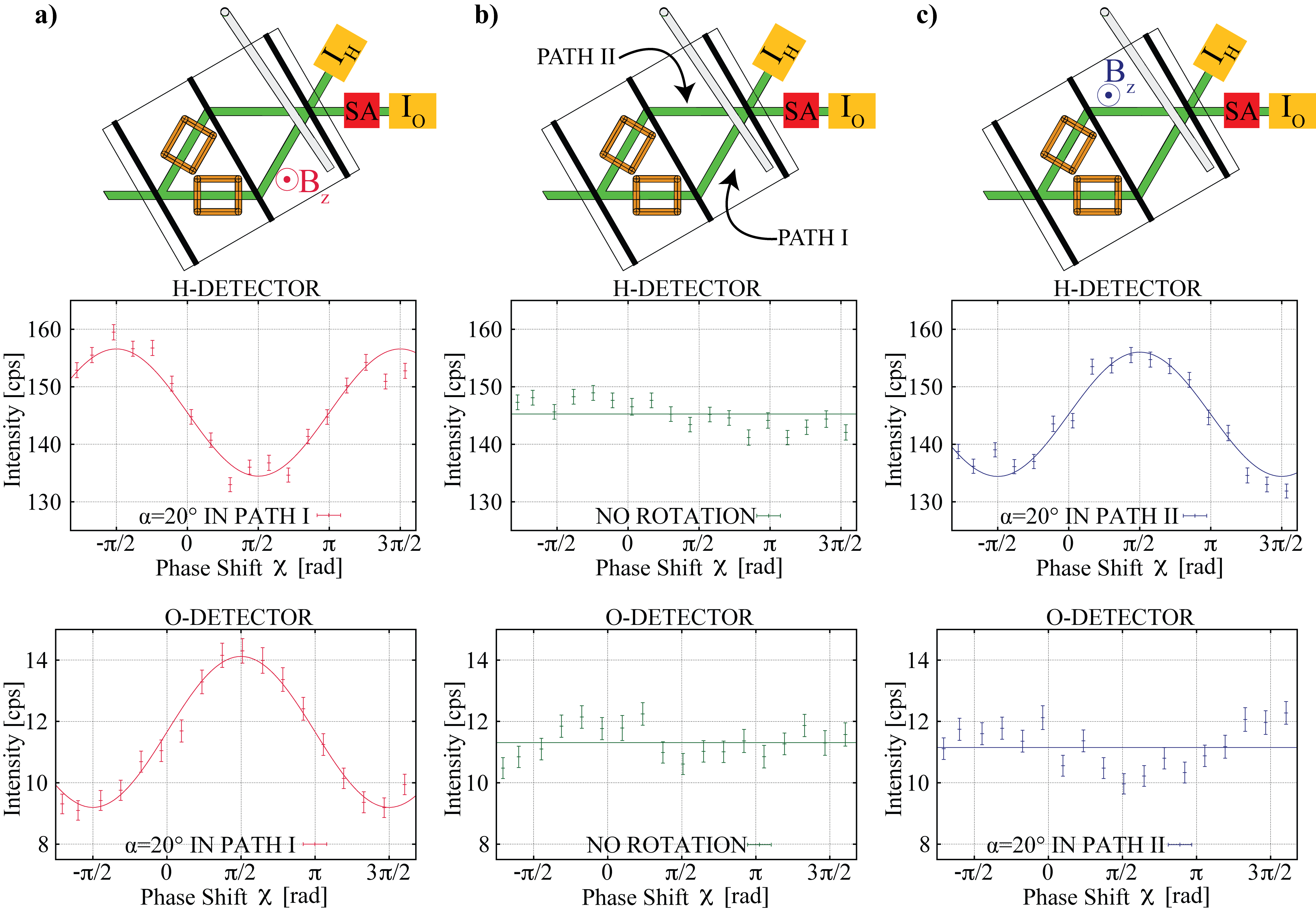}
\caption{Measurement of $\langle \hat{\sigma}_z\hat{\Pi}_{{I}}\rangle_w$ and $\langle \hat{\sigma}_z\hat{\Pi}_{{II}}\rangle_w$ applying small additional magnetic fields. The intensity of the O-beam (with the spin-analysis) and the H-beam (without the spin analysis) is plotted as a function of the relative phase $\chi$. \textbf{a)} A magnetic field in path $I$: Interference fringes appear both at the postselected O-detector and the H-detector. \textbf{b)}A reference measurement without any additional magnetic fields. Since the spin states inside the interferometer are orthogonal, interference fringes appear neither in the O-, nor the H-detector.  \textbf{c)}A magnetic field in path $II$: No interference fringes can be seen at the spin postselected O-detector, whereas a clear sinusoidal intensity modulation is visible at the H-detector without spin analysis. The measurements suggest that for the successfully postselected ensemble (only the O-detector) the neutrons' spin travels along path $I$.}\label{fig:cat_spin}
\end{figure}

A fundamental difference between classical and quantum physics is, that the initial and final boundary conditions of a quantum system can be selected independently \cite{ABL64}. It was Aharonov, Albert and Vaidman~\cite{aav88} who first introduced the weak value defined as 
\begin{equation}
\langle\hat{A}\rangle_w={{ {\langle \psi _{f}\! \mid \hat{A} \mid \!\psi _{i} \rangle} \over {\langle \psi _{f} \!\mid \!\psi _{i} \rangle}}}
\label{expweak}
\end{equation}
where $|\psi_{i}\rangle$ and $|\psi_{f}\rangle$ are the initial (``preselected'') and final (``postselected'') states of the system and $\hat{A}$ is an observable of the system being measured. $\langle\hat{A}\rangle_w$ represents information obtained by weakly coupling the system to a measurement device, i.e. a probe. Such ``weak measurements" allow information to be obtained with minimal disturbance~\cite{tollaksen07}.\\
A surprising effect originating from pre- and postselection of a system is the ability to ``separate" the location of a system from one of its properties~\cite{tollaksen01,aharonov05,aharonov13,matzkin13}, as suggested by the Cheshire Cat story: ``Well! I've often seen a cat without a grin," thought Alice; ``but a grin without a cat! It's the most curious thing I ever saw in all my life!"~\cite{carroll}.\\
The essential property of a \textit{quantum} Cheshire Cat in an interferometer is, that the cat itself is located in one beam path, while its grin is located in the other one \cite{aharonov13}. An artistic depiction of this behaviour is shown in Fig. \ref{fig:cat}.
 In our experiment, the neutron plays the role of the cat and the cat's grin is represented by the neutron's spin. The system is initially prepared so that after entering the beamsplitter its quantum state is given by
\begin{align}
|\psi_{i}\rangle=\frac{1}{\sqrt{2}}|S_x;+\rangle |I\rangle+\frac{1}{\sqrt{2}}|S_x;-\rangle |II\rangle,\label{eqn:initialstate}
\end{align}
where $|I\rangle$ ($|II\rangle$) stands for the spatial part of the wavefunction along path $I$ (path $II$) of the interferometer and $|S_x;\pm\rangle$ denotes the spin state in $\pm\hat{x}$ direction. In order to observe the quantum Cheshire Cat, after we preselect the ensemble, we will next perform weak measurements on both the  neutrons' population in a given path along with the value of the spin in a given path.  After these weak measurements, the ensemble is then postselected in the final state:
\begin{align}
|\psi_{f}\rangle=\frac{1}{\sqrt{2}}|S_x;-\rangle\biggl[ |I\rangle+|II\rangle\biggr].\label{eqn:finalstate}
\end{align}
Weak measurements allow to obtain information on the value of an observable without destroying the subsequent evolution of the system. Therefore they yield the information about the location of the neutrons and their spin in the context of pre- and postselection.\\ 
We can calculate the weak values of the projection operators on the neutron path eigenstates $\hat{\Pi}_{j}\equiv|{j}\rangle\langle{j}|$, with $j=I,II$. From equations (\ref{expweak}) to (\ref{eqn:finalstate}) we obtain $\langle\hat{\Pi}_{I}\rangle_{w}=0$ and $\langle\hat{\Pi}_{II}\rangle_{w}=1$. Theses values tell us the neutrons' population in path $I$ and $II$ respectively.\\
Now we ask about the weak value of the location of the neutrons' spin. It would be natural to assume that this is just the product of the weak values, namely $\langle\hat\sigma_z\rangle_w\langle\hat{\Pi}_{II}\rangle_{w}$, i.e. the spin is ``located" where the spatial part of the wavefunction is located on path $II$.  But this is false. The reason for this is that in general, the weak value of a product of observables is not equal to the product of their weak values (even if the observables commute).  The appropriate observable to ascertain the weak value of the neutrons' spin on path $II$ is therefore $\langle\hat\sigma_z\hat{\Pi}_{II}\rangle_{w}$ which is not equal to $\langle\hat\sigma_z\rangle_w\langle\hat{\Pi}_{II}\rangle_{w}$. In fact, we find $\langle \hat{\sigma}_z\hat{\Pi}_{I}\rangle_w=1$ and $\langle \hat{\sigma}_z\hat{\Pi}_{II}\rangle_w=0$.  Alice would say ``Curiouser and curiouser."\\
The experiment presented here was performed at the S18 interferometer beam line at the research reactor at the Institut Laue Langevin~\cite{S18}. The setup is shown in Fig. \ref{fig:setup}.
A monochromatic neutron beam with a wave length of $\lambda=1.92$ \AA~passes magnetic birefringent prisms, which polarize the neutron beam. To avoid depolarization a magnetic guide field pointing in the $+\hat{z}$ direction is applied around the whole setup. A spin turner, rotates the neutron spin by $\pi/2$ into the xy plane. The neutron's spin wave function is then given by $|S_x;+\rangle$. Subsequently the neutrons enter a triple-Laue interferometer \cite{rauch00,hasegawa11}. Inside the interferometer a spin rotator in each beam path, allows the generation of the preselected state $|\psi_{\text{i}}\rangle$.\\
A phase shifter is inserted into the interferometer to tune the relative phase $\chi$ between path $I$ and path $II$. Hence, a general postselected path state is given by $\frac{1}{\sqrt2}\left(|I\rangle+\mathrm e^{-i\chi}|II\rangle\right)$. Of the two outgoing beams of the interferometer only the O-beam is affected by a spin analysis. The H-beam without a spin analysis is used as a reference monitor for phase and count rate stability. The spin postselection of the O-beam is done using a spin turner and a polarizing supermirror. Both outgoing beams are measured using $^3\text{He}$ detectors with very high efficiency (over 99\%). All measurements presented here are performed in a similar manner: The phase shifter is rotated, thereby scanning $\chi$ and recording interferograms. The interferograms allow to extract the intensity for $\chi=0$. This ensures that the path postselection is indeed carried out on the state $\frac{1}{\sqrt2}\left(|I\rangle+|II\rangle\right)$ corresponding to the the O-beam.\\
To determine the neutrons' population in the interferometer's paths $\langle\hat{\Pi}_{j}\rangle_w$ are determined by inserting absorbers into the respective path $j$ of the interferometer.$\langle\hat{\Pi}_{j}\rangle_w$ is evaluated in three steps: First a reference intensity is measured by performing a phase shifter scan of the empty interferometer to determine $\text{I}^\text{REF}$. For the reference measurement the spin states inside the interferometer are orthogonal. Therefore the interferogram shows no intensity oscillation. As a second step, an absorber with known transmissivity of $T=0.79(1)$ is inserted into path $I$ and the phase shifter scan is repeated, which yields $\text{I}_{I}^\text{ABS}$. Finally the absorber is taken out of path $I$ and put into path $II$. The subsequent phase shifter scan allows the extraction of $\text{I}_{II}^\text{ABS}$. A typical measurement result is depicted in Fig. \ref{fig:setup_allabs_meas}. In path $I$ the absorber has no effect. In comparison to the reference intensity no significant change can be detected in the count rate. As opposed to this, the very same absorber decreases the intensity, when it is put in path $II$.
This already tells us that the neutrons' population in the interferometer is obviously higher in path $II$ than it is in path $I$.\\
The weak measurements of the neutrons's spin in each path are achieved by applying additional magnetic fields in one or the other beam path. This causes a small spin rotation which allows to probe the presence of the neutron's magnetic moment in the respective path. If there is a magnetic moment present in the path, the field has an effect on the measured interference fringes. If no change in the interferogram can be detected, there is no magnetic moment present in the path, where the additional field is applied.
The condition of a weak measurement is fulfilled, by tuning the magnetic field small enough. A spin rotation of $\alpha=20^\circ$ is applied, corresponding to a wave function overlap of 98.5 \%. The results of the measurement procedure are shown in Fig. \ref{fig:cat_spin}, where they are compared to the reference measurement performed with the empty interferometer. Due to the orthogonal spin states inside the interferometer the interferogram shows no intensity oscillation for both O- and H-detector. An additional magnetic field in path $I$ leads to the appearance of interference fringes at both O- and H-beam, giving $\text{I}_{I}^\text{MAG}$. Obviously a magnetic moment is present in path $I$, since the interferogram changes. Now the same field is applied in in path $II$, to obtain $\text{I}_{II}^\text{MAG}$. The field induces no intensity modulation for the spin postselected O-beam, while a sinusoidal oscillation appears at the H-beam, which has no spin analysis. For the successfully postselected ensemble the neutron's spin only travels along path $I$.
The recorded interferograms allow to extract the intensities. For the respective measurements they are given by (all in counts per second): $\text{I}^\text{REF}=11.25(5)$, $\text{I}_{I}^\text{ABS}=10.90(9)$, $\text{I}_{II}^\text{ABS}=8.83(8)$, $\text{I}_{I}^\text{MAG}=11.59(6)$ and $\text{I}_{II}^\text{MAG}=10.97(6)$. From these values we obtain the results of the weak measurements (see Methods for details). The error in the determination of $\langle \hat{\sigma}_z\hat{\Pi}_{j}\rangle_w$ is dominated by the statistical error of the count rate. To gain precision the measurements were repeated several times. Table \ref{tab:results} summarizes the final results.
\begin{table}[ht!]
\centering
\begin{tabular}{ccc}
&  \multicolumn{1}{ c }{Path $I$} & \multicolumn{1}{ c }{Path $II$}\\ \hline \hline
\multicolumn{1}{ c }{$\langle \hat{\Pi}_j\rangle_w$} &$\ 0.139\pm0.041\ $ &$\ 0.960\pm 0.058\  $ \\ \cline{1-3} 
\multicolumn{1}{ c }{$|\langle \hat{\sigma}_z\hat{\Pi}_j\rangle_w|^2$} & $\  0.999\pm0.252 \ $ & $\ 0.172\pm0.223 \ $ \\ \cline{1-3}
\end{tabular}
\caption{Summary of the results}\label{tab:results}
\end{table}\\
Theory predicts that for the pre- and postselected states $|\psi_i\rangle$ and $|\psi_f\rangle$ the weak values for the spin in paths $I$ and $II$ are given by $|\langle \hat{\sigma}_z\hat{\Pi}_I\rangle_w|^2=1$ and $|\langle \hat{\sigma}_z\hat{\Pi}_{II}\rangle_w|^2=0$, while the weak values for the neutrons' population along these paths are $\langle \hat{\Pi}_{I}\rangle_w=0$ and $\langle \hat{\Pi}_{II}\rangle_w=1$. Within the error the experiment fully confirms this prediction.\\
In our neutron interferometer experiment a quantum Cheshire cat is created by applying an appropriate pre- and postselection on non-separable and separable spin path states. In order to infer the neutrons' population and the location of their spin, weak measurements of the operators $\hat{\Pi}_j$ and $\hat{\sigma}_z\hat{\Pi}_j$ are carried out. The obtained results exhibit the characteristics of a quantum Cheshire Cat in an interferometer: We observe that an absorber in path $I$ does not change the measurement outcome, while a magnetic field does. In contrast to that the absorber has an effect in path $II$, while the magnetic field has none. The neutrons behave as if particle and magnetic property are spatially separated while travelling through the interferometer. 

\section*{Methods}

\subsection*{A) Weak Measurement of $\langle\hat{\Pi}_{I}\rangle_w$ and $\langle\hat{\Pi}_{II}\rangle_w$}

The weak measurement of $\langle\hat{\Pi}_{{I}}\rangle_{w}$ and $\langle\hat{\Pi}_{{II}}\rangle_{w}$ is performed using absorbers with a high transmissivity (i.e., a ``weak absorption"). Phenomenologically, an absorption in path $j$ can be represented by an imaginary optical potential \begin{align}
\hat{V}_{j}=-i\mu_{j}(r)\hat\Pi_{j},
\end{align} 
where the absorption coefficient is given by $\text{M}_j=\int \mu_j(r)\mathrm dr$ in path $j$ with $r$ is integrated on the absorber slab thickness. For weak absorption $\text{M}_{j}$ can be related to the transmissivity $T_{j}$ through $\text{M}_{j}\approx1-\sqrt{T_{j}}$ \cite{sears89}. Omitting for simplicity the free evolution operators in the following expressions, the wavefunction after the wavepacket has interacted with the absorber along path $j$ is
\begin{align}
|\psi'\rangle=\mathrm e^{-i\int\mathrm dr \hat{V}_{j}}|\psi_{i}\rangle\approx\left\lbrack1-\text{M}_{j}\hat\Pi_{j}+\cdots\right\rbrack|\psi_{i}\rangle
\end{align} 
for small $\text{M}_{j}$. Using $\langle\hat{\Pi}_{j}\rangle_{w}=\frac{\langle\psi_{f}|\hat{\Pi}_{j}|\psi_{i}\rangle}{\langle\psi_{f}|\psi_{i}\rangle}$ the intensity for the postselected outcome takes the form 
\begin{equation}
\text{I}_{\text{j}}^{\text{ABS}}=|\langle\psi_{f}|\psi_{i}\rangle|^{2}\left\lbrack1-2\text{M}_j\langle\hat{\Pi}_{{j}}\rangle_{w} \right\rbrack
\end{equation}
where contribution of the imaginary part of $\langle\hat{\Pi}_{j}\rangle_{w}$ is treated as being of order or less then $\text{M}^2$: This is justified by the contrast $C\le0.024(5)$ of the empty interferometer measurements with the experimental pre- and postselected states $|\psi_i\rangle$ and $|\psi_f\rangle$. Since $\text{I}^{\text{REF}}=|\langle\psi_{f}|\psi_{i}\rangle|^{2}$, the weak values can be extracted from the observation of $\text{I}_{{j}}^{\text{ABS}}/\text{I}^{\text{REF}}$.

\subsection*{B) Weak measurement of $\langle \hat{\sigma}_z\hat{\Pi}_{{I}}\rangle_w$ and $\langle \hat{\sigma}_z\hat{\Pi}_{{II}}\rangle_w$}
 
The weak value of $\hat{\sigma}_z\hat{\Pi}_{{j}}$ is determined using path conditioned spin rotations. To measure $\langle\hat{\sigma}_z\hat{\Pi}_{{j}}\rangle_w$, a small magnetic field is applied in path $j$. The interaction Hamiltonian for this measurement is
\begin{align}
\hat{H}_{j}=-\gamma\frac{\hat{\sigma}_z}{2}B_z\hat{\Pi}_j
\end{align}
where $\gamma$ is the gyromagnetic ratio and $B_z$ an externally applied magnetic field. $\hat{\Pi}_{j}$ denotes the fact the the magnetic field is applied only in the region along path $j$. $\vec{\sigma}$ is the Pauli vector; applying a magnetic field along $\hat{z}$ leads to the $\hat{\sigma}_{z}$ component of the Pauli matrix, which is the generator of rotations around the z-axis. Thus a small rotation around $z$ on path $j$ is generated through the coupling between the magnetic field and the spin projection of the neutron on the $z$ axis.
The rotation angle produced by this Larmor precession effect will be labeled by $\alpha$; its magnitude is proportional to the magnetic field strength \cite{rauch00}. The evolution of the initial state caused by the weak measurement is given by
\begin{align}
|\psi'\rangle=\mathrm e^{-i\int\mathrm dt \hat{H}_{j}}|\psi_i\rangle=\mathrm e^{i\alpha\hat{\sigma}_z\hat{\Pi}_{{j}}/2}|\psi_i\rangle\approx\left\lbrack1+\frac{i\alpha}{2}\hat{\sigma}_z\hat{\Pi}_{{j}}+\cdots\right\rbrack|\psi_i\rangle
\end{align}
After postselection for the outcomes corresponding to the final state $|\psi_f\rangle$, the intensity at the O-detector is
\begin{align}
\text{I}_{j}^{\text{MAG}}&=|\langle \psi_f|\psi'\rangle|^2=|\langle \psi_f|\psi_i\rangle|^2-\frac{\alpha^2}{4}\langle \psi_f|\hat{\Pi}_{{j}}|\psi_i\rangle\langle \psi_i|\psi_f\rangle+\frac{\alpha^2}{4}|\langle \psi_f|\hat{\sigma}_z\hat{\Pi}_{{j}}|\psi_i\rangle|^2,\label{eqn:intpathj}
\end{align}
taking into account $\alpha$ up to $\alpha^2$. Again the contribution of the imaginary part of $\langle\hat{\sigma}_z\hat{\Pi}_{{j}}\rangle_w$ is treated as being of order or less then $\alpha^3$: As in Methods A, this can be justified by the contrast of the empty interferometer. The contributions of $\langle \psi_f|\hat{\Pi}_{{j}}|\psi_i\rangle$ are determined experimentally as well. Taking them into account yields:
\begin{subequations}
\begin{eqnarray}\label{eq:Bell-angles}
\text{I}_{I}^{\text{MAG}}=|\langle \psi_f|\psi_i\rangle|^2\left\lbrack1+\frac{\alpha^2}{4}|\langle\hat{\sigma}_z\hat{\Pi}_{{I}}\rangle_w|^2\right\rbrack\\
\text{I}_{II}^{\text{MAG}}=|\langle \psi_f|\psi_i\rangle|^2\left\lbrack1-\frac{\alpha^2}{4}+\frac{\alpha^2}{4}|\langle\hat{\sigma}_z\hat{\Pi}_{{II}}\rangle_w|^2\right\rbrack
\end{eqnarray}
\end{subequations}
Since $\text{I}^{\text{REF}}=|\langle\psi_{f}|\psi_{i}\rangle|^{2}$, the weak values are extracted from the measurements of the intensities with the magnetic field along path $I$, along path $II$ and with the magnetic field turned off.

\begin{acknowledgments}

We acknowledge support by the Austrian Science Fund (FWF), as well as from the Austrian-French binational Amadeus Project. We also acknowledge Leon Filter's artistic depiction of the Cheshire Cat.\\

\textbf{Author Information} 
Correspondence and requests for materials should be addressed to Y.H. (Hasegawa@ati.ac.at).\\
 
\end{acknowledgments}

\end{document}